\documentclass[letterpaper, 10 pt, conference]{ieeeconf}
\pdfoutput=1
\IEEEoverridecommandlockouts                              

\usepackage{graphics} 
\usepackage{epsfig} 
\usepackage{times} 
\usepackage{amsmath} 
\usepackage{amssymb}  
\usepackage{subcaption}
\usepackage{xcolor}
\usepackage[hidelinks]{hyperref}

\title{\LARGE \bf
Online Phase Estimation of \\ Human Oscillatory Motions using Deep Learning 
}

\author{Antonio Grotta$^{1}$, Francesco De Lellis$^{2}$
    \thanks{$^{1}$Scuola Superiore Meridionale, Italy}
    \thanks{$^{2}$Department of Electrical Engineering and Information Technologies, University of Naples
    Federico II, Italy}
    \thanks{This work was supported by the EU Research Project SHARESPACE (EU HORIZON-CL4-2022-HUMAN-01-14). For more info see {\tt www.sharespace.eu}}
}

\begin{document}

\maketitle
\thispagestyle{empty}
\pagestyle{empty}

\begin{abstract}

Accurately estimating the phase of oscillatory systems is essential for analyzing cyclic activities such as repetitive gestures in human motion. 
In this work we introduce a learning-based approach for online phase estimation in three-dimensional motion trajectories, using a Long Short-Term Memory (LSTM) network. 
A calibration procedure is applied to standardize trajectory position and orientation, ensuring invariance to spatial variations. 
The proposed model is evaluated on motion capture data and further tested in a dynamical system, where the estimated phase is used as input to a  reinforcement learning (RL)-based control to assess its impact on the synchronization of a network of Kuramoto oscillators.

\end{abstract}

\section{Introduction}
Phase estimation is a critical aspect of oscillatory system analysis, with wide-ranging applications in human movement science, robotics, neuroscience, and biomechanics \cite{pikovsky2001synchronization, rosenblum2021real}. Accurately estimating the phase of cyclic movements is essential for understanding coordination dynamics, motor control, and movement synchronization \cite{lee2021lstm, alderisio2017a, lombardi2019deeplearning, zhai2016virtual}.

Traditional signal processing techniques, such as the Hilbert transform, are commonly used to extract phase information from movement signals \cite{pikovsky2001synchronization, rosenblum1996phase}. However, such method typically requires access to the entire signal, rendering it unsuitable for scenarios where phase must be estimated online as new data are collected. This limitation is particularly pronounced when dealing with multi-dimensional motion signals, where standard phase extraction methods often struggle with complex, nonlinear oscillatory patterns.

Several online phase estimation methods have been proposed, primarily focusing on one-dimensional (1D) oscillatory signals, such as neural activity, tremors, and simple limb movements. For instance, studies have developed real-time techniques for neural oscillation detection \cite{Chen2013RealTime}, phase-locked tremor suppression \cite{Schreglmann2021NonInvasive}, and estimation of Parkinsonian patients' beta-band brain activity \cite{rosenblum2021real}.

While effective for 1D signals, these approaches do not readily generalize to multi-dimensional human motion, where multiple oscillatory components interact across different body segments. Other studies have explored online phase estimation in three-dimensional (3D) motion, particularly within gait analysis. Here, machine learning models predict continuous gait phase using data from wearable sensors. Various deep learning approaches have been tested for this task, often leveraging sensor fusion techniques to integrate data from multiple sensor types. Examples include the use of recurrent neural networks \cite{seo2019rnn}, convolutional neural networks \cite{MOURACOELHO2022117306}, and hybrid architectures combining physics-based modeling with learning-based inference \cite{xu2020adaptive}. As a specific instance, in \cite{lee2021lstm}, the authors proposed a Long Short-
Term Memory (LSTM)-based approach for continuous gait phase estimation tailored for robotic transfemoral prostheses.

However, these gait-focused methods are inherently task-specific. They rely on biomechanical signals and kinematic features uniquely characteristic of human walking patterns---such as ground reaction forces, heel-strike, and toe-off events---limiting their applicability to other types of 3D oscillatory motion.

In this work, we propose a novel framework for online phase estimation of general 3D oscillatory motion, addressing the gap between existing online 1D phase estimation and gait-specific 3D methods. We first extract phase signals offline using principal component analysis (PCA) and the Hilbert transform applied to fully executed end-effector motions. To enable online inference, we then use an LSTM model trained to predict phase using only a limitedd window of past observations.

The proposed model is validated both in standard supervised learning settings and in combination with a pre-trained reinforcement learning (RL) agent. Specifically, the trained LSTM is used to estimate a phase signal from 3D motions whose true phase is computed by coupled Kuramoto oscillators. The final performance is evaluated to quantify how small deviations introduced by the LSTM model affect the pre-trained RL-based control system \cite{GROTTA202437}. Our results indicate that our model provides low estimation errors, allowing the RL agent to maintain synchronization properties and demonstrating its effectiveness in practical applications.

\section{Problem Formulation}

\label{sec:problem_formulation}
Let $\mathcal{D}_{\text{tot}} = \{\mathbf{X}^{i}\}_{i=1}^{N}$ represent a dataset comprising $N$ human 3D motion trajectories. Each trajectory $\mathbf{X}^{i} = [\mathbf{x}^{i}_1,\, \mathbf{x}^{i}_2,\, \dots,\, \mathbf{x}^{i}_{T_i}]^\top \in \mathbb{R}^{T_i\times 3}$ is a time series of length $T_i$, where each $\mathbf{x}^{i}_t \in \mathbb{R}^3$ denotes the 3D spatial position at time index $t$. We assume the underlying motion exhibits cyclical patterns characterized by a time-varying phase $\theta_t^{i} \in [-\pi, \pi]$, which quantifies the continuous progress through the movement cycle.

Our objective is to learn a function $\mathcal{F}_{\boldsymbol{\psi}}: \mathbb{R}^{w\times 3} \to [-\pi, \pi]$, parameterized by $\boldsymbol{\psi}$, that estimates the phase $\widehat{\theta}_t^{i}$ at time $t$ using only a short sub-trajectory $\mathbf{x}^{i}_{t-w+1 : t} = [\mathbf{x}^{i}_{t-w+1},\,\dots,\, \mathbf{x}^{i}_{t}]^\top$ of fixed length $w$. This is expressed as:
\begin{equation} \label{eq:approximator_function}
\widehat{\theta}_t^{i} = \mathcal{F}_{\boldsymbol{\psi}}\big(\mathbf{x}^{i}_{t-w+1 : t}\big),
\end{equation}
for any valid index $t$ within trajectory $i$. The key advantage is the model's dependence solely on a limited window of the 3D signal. Since true phase estimation relies on the temporal evolution and past observations, we employ an LSTM model for this approximator. The training minimizes the mean squared error between the LSTM output and \emph{ground-truth} phase labels \cite{hochreiter1997long}, which are computed by applying Principal Component Analysis (PCA) followed by a Hilbert transform to the acquired 3D motion data \cite{bracewell2000fourier}.

\section{Data Acquisition} \label{sec:data_acquisition}
To train and evaluate our phase estimation model, we collected a dedicated dataset of 3D human motion trajectories with clear cyclical patterns. The experimental protocol involved six groups, each with four participants performing oscillatory finger movements along a specific axis. Each group repeated the experiment six times under two conditions:
\begin{itemize}
    \item \emph{Blinded condition}: Participants performed movements without visual feedback of others, ensuring self-generated motion.
    \item \emph{Coupled condition}: Participants performed movements while seeing each other, allowing for potential interpersonal synchronization.
\end{itemize}
This protocol yielded a rich dataset capturing diverse human behaviors and variability, including \emph{intra-individual} fluctuations across trials and \emph{inter-individual} differences arising from social interaction and synchronization \cite{alderisio2017a}.

\begin{figure}[t]
    \centering
    \subfloat[]
    {
        \includegraphics[scale=0.34]{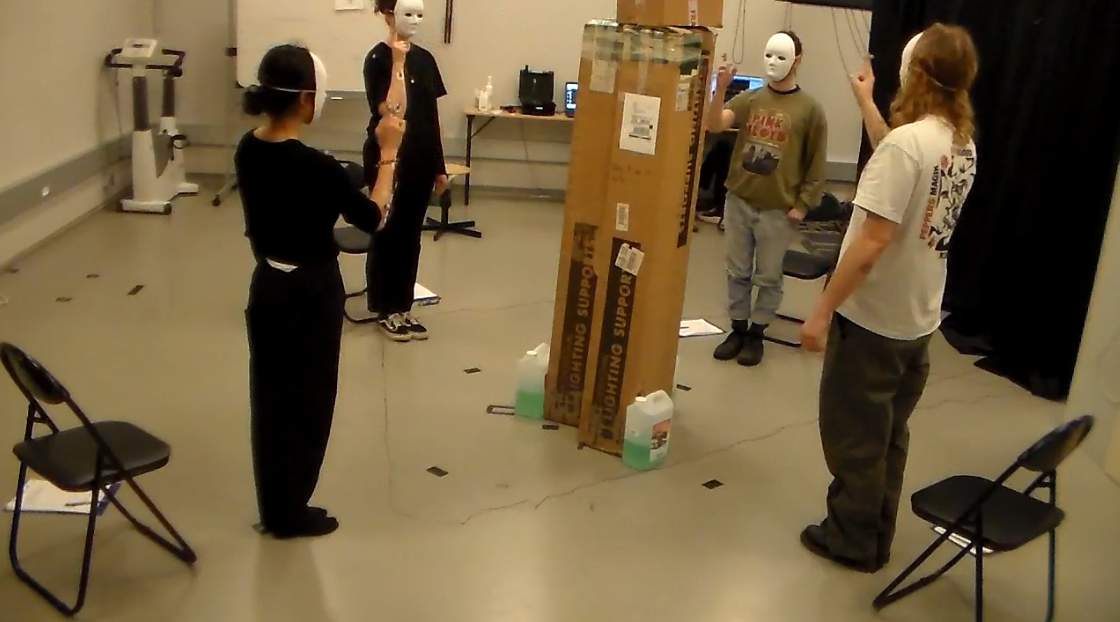}
        \label{fig:group_interaction_real}
    }
    
    \subfloat[]    
    {
        \includegraphics[scale=0.26]{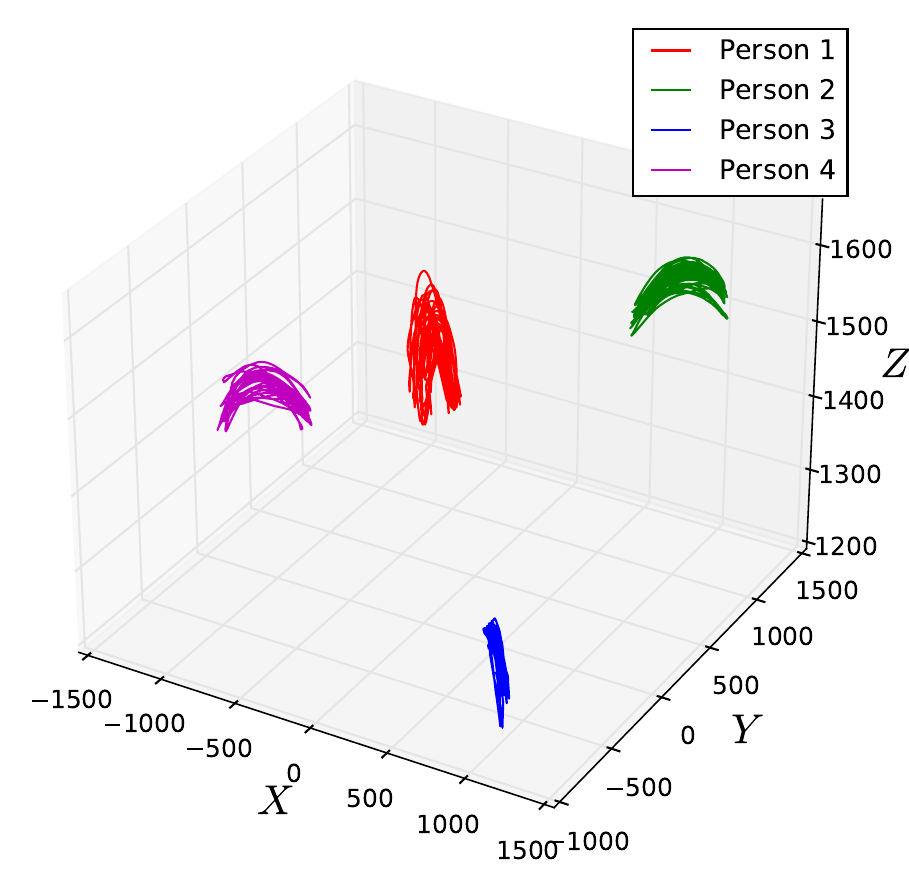}
        \label{fig:group_interaction_3d}
    }
    \subfloat[]    
    {
        \includegraphics[scale=0.26]{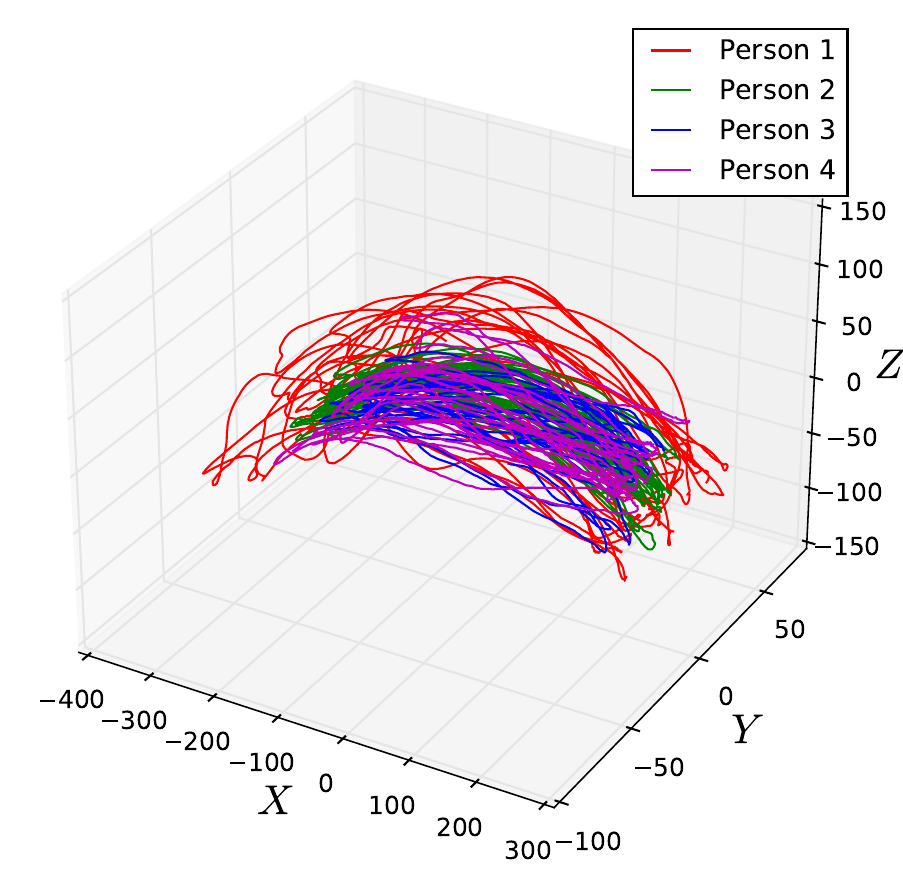}
        \label{fig:group_interaction_3d_calibrated}
    }
    \caption{Visualization of the experiment, with a screenshot of the setup (a), the corresponding three-dimensional motion trajectories of the participants (b) and the same trajectories after calibration (c).}
    \label{fig:group_interaction}
\end{figure}

Motion data were recorded using a high-precision \emph{Vicon Motion Systems Ltd} motion capture system, capturing three-dimensional end-effector movements at 100 Hz. Fig.~\ref{fig:group_interaction_real} shows the experimental setup. Figs.~\ref{fig:group_interaction_3d} and \ref{fig:group_interaction_3d_calibrated} illustrate the raw and calibrated data, respectively, with calibration details provided in Sec.~\ref{sec:motion_data_representation}.

\section{Motion Data Calibration}
\label{sec:motion_data_representation}
To accurately estimate motion phase representations, it is crucial to account for how spatial variations affect trajectory data. Differences in sensor placement, object orientation, and viewpoint can lead to inconsistencies in trajectory representations, rendering direct phase estimation unreliable.

To overcome this issue, we apply a calibration procedure to the training dataset, where trajectories were collected in a global reference frame. This process removes positional offsets and aligns orientations, ensuring that trajectory representations become invariant to the specific recording setup (sensor placement and orientation) while preserving the underlying motion dynamics.

This calibration step is essential for training the LSTM to obtain the phase estimator $\mathcal{F}_{\boldsymbol{\psi}}$ (defined in Eq.~\eqref{eq:approximator_function}) that is independent of the recording setup used during data collection. During real-time deployment, we assume that trajectory data is directly collected in a local frame attached to each human participant, thereby removing the need for additional calibration at inference time.

The calibration procedure consists of two main steps: (i) centering the trajectory to remove positional offsets, and (ii) alignment via rotation matrix to standardize orientation.

\paragraph{Centering}
A key step in preprocessing is the elimination of absolute positional offsets so that each trajectory is analyzed relative to its own center, independent of its original spatial placement. This makes the movement pattern the primary focus and allows our learning model to concentrate solely on these patterns, irrespective of the trajectory's absolute position or its position relative to others.

To achieve position invariance, for each trajectory $\mathbf{X}^{i}$ in the dataset $\mathcal{D}_{\text{tot}}$, we first compute its mean 3D position:
\begin{equation}
\bar{\mathbf{x}}^{i} = \frac{1}{T_i} \sum_{t=1}^{T_i} \mathbf{x}^{i}_t,
\label{eq:mean_calibration}
\end{equation}
where \( \bar{\mathbf{x}}^{i} = [\bar{x}^{i}, \bar{y}^{i}, \bar{z}^{i}]^\top \in \mathbb{R}^3 \). The entire trajectory is then shifted such that its mean coincides with the origin:
\begin{equation}
\mathbf{X}'^{i} = \mathbf{X}^{i} - \bar{\mathbf{x}}^{i}.
\label{eq:shifted_traj}
\end{equation}
This operation effectively removes the absolute location dependence while preserving the relative shape and motion dynamics.

\paragraph{Alignment}
After centering, trajectories may still exhibit variations in orientation due to differences in sensor placement or recording perspective. 
Unlike centering, which is a local transformation, alignment is a global calibration step, as it ensures that all trajectories share the same primary direction.
Given a centered trajectory $\mathbf{X'}^{i}$, a single principal direction is extracted via PCA, producing a unit vector $\mathbf{p}^{i} =[p_1^{i}, p_2^{i}, p_3^{i}]\in \mathbb{R}^3$. 
To enforce a consistent orientation across trajectories, we align $\mathbf{p}^{i}$ with an arbitrary reference axis, say the $x$-axis $\mathbf{e}_x = [1,0,0]^\top$.
The rotation matrix $\mathbf{R}^{i}\in \mathbb{R}^{3\times 3}$ is computed using Rodrigues' formula \cite{Dai2015EulerRodrigues}
\begin{equation} 
    \mathbf{R}^{i} = \mathbf{I} + \mathbf{V}_\times^{i} + \mathbf{V}_\times^{i2} \frac{1 - c^{i}}{(s^{i})^2},
    \label{eq:rotation_calibration}
\end{equation} 
where $\mathbf{I} \in \mathbb{R}^{3 \times 3}$ is the identity matrix, and $\mathbf{V}_\times^{i}\in \mathbb{R}^{3 \times 3}$ is the skew-symmetric matrix computed as
\begin{equation}
    \mathbf{V}_\times^{i} = \mathbf{p}^{i} \times \mathbf{e}_x = \begin{bmatrix}
        0 & -p_3^{i} & p_2^{i} \\
        p_3^{i} & 0 & -p_1^{i} \\
        -p_2^{i} & p_1^{i} & 0
    \end{bmatrix}.
\end{equation}
The scalar terms $c^{i}$ and $s^{i}$ are defined as
\begin{equation}
    c^{i} = \mathbf{p}^{i} \cdot \mathbf{e}_x, \quad
    s^{i} = \|\mathbf{p}^{i}\|.
\end{equation}
The aligned trajectory is obtained by applying $\mathbf{R}^{i}$ to the shifted trajectory in Eq.~\eqref{eq:shifted_traj}:
\begin{equation}
    \tilde{\mathbf{X}}^{i} = \mathbf{X}'^{i} \mathbf{R}^{i}.
    \label{eq:rot_traj}
\end{equation}
If $s^{i} \approx 0$, meaning that $\mathbf{p}^{i}$ is already aligned with $\mathbf{e}_x$, no transformation is applied, and $\mathbf{R}^{i} = \mathbf{I}$.
Figure \ref{fig:group_interaction_3d_calibrated} illustrates an example of the calibration applied to the raw trajectories shown in Figure \ref{fig:group_interaction_3d}.
As we can see, the trajectories are all centered at zero, as each one has been adjusted by subtracting its mean position. 
Additionally, they are rotated with respect to the $x$-axis to standardize orientation.

\section{Training and Validation} \label{sec:training}
Since phase estimation inherently depends on temporal motion data, we employ an LSTM network architecture \cite{hochreiter1997long}. Our model processes a window of $w = 10$ past samples of 3D end-effector position and velocity. Each sample consists of a 3D position vector and a 3D velocity vector, resulting in $6$ input features. 
The resulting input tensor has shape $10 \times 6=60$, which is fed into a single-layer LSTM with 128 hidden units. The output is then passed through a fully connected layer with 2 linear neurons representing the sine and cosine of the estimated phase. Additionally, a dropout layer with a rate of 0.2 is applied after the hidden layer to mitigate overfitting \cite{srivastava2014dropout}. This neural architecture allows us to effectively represent the temporal dependencies described in Eq.~\eqref{eq:approximator_function}.

\subsection{Training}
To train our model, we pool together all data obtained from both the \emph{Blinded} and \emph{Coupled} experimental conditions, as described in Section~\ref{sec:data_acquisition}. We then compute the ground-truth phase labels associated with the dataset by applying PCA to extract the dominant direction of motion, followed by the Hilbert transform to obtain the offline phase labels, denoted collectively as $\boldsymbol{\Theta}$.

To effectively assess the model's generalization capabilities, we partition the total dataset \( \mathcal{D}_{\text{tot}} \) into distinct training, validation, and test subsets. This partition serves three main purposes: (i) selecting specific groups whose data are used for training, (ii) evaluating the model's ability to generalize phase estimation across subjects within the training/validation groups using the validation dataset, and (iii) testing generalization across different human subjects using the test dataset (groups not seen during training/validation).

Specifically, the training and validation data, denoted together as \( \mathcal{D} \), originate from groups 1, 2, and 3. Within \( \mathcal{D} \), we allocate 70\% of the trajectories for training ($ \mathcal{D}_{\text{train}} $) and the remaining 30\% for validation ($ \mathcal{D}_{\text{val}} $). Formally:
\begin{equation}
\mathcal{D}_{\text{train}} = \{\mathbf{X}^{i} \mid i \in \mathcal{I}_{\text{train}}\}, \quad
\mathcal{D}_{\text{val}} = \{\mathbf{X}^{i} \mid i \in \mathcal{I}_{\text{val}}\},
\end{equation}
where \( \mathcal{I}_{\text{train}} \) and \( \mathcal{I}_{\text{val}} \) are disjoint index sets such that $|\mathcal{I}_{\text{train}}| = 0.7 L$ and $|\mathcal{I}_{\text{val}}| = 0.3 L$, with \( L \) being the total number of trajectories in \( \mathcal{D} \).

To ensure numerical stability and consistent input scaling, we apply min-max normalization \cite{duda2001pattern}, rescaling each trajectory's dimensions into the fixed range \([0,1]\). The scaling parameters (minimum and maximum values) used for normalization are computed exclusively from the training and validation data combined (\( \mathcal{D}_{\text{train}} \cup \mathcal{D}_{\text{val}} \)).

The test dataset \( \mathcal{D}_{\text{test}} \) consists of data from groups 4, 5, and 6:
\begin{equation}
\mathcal{D}_{\text{test}} = \{\mathbf{X}^{i} \mid i \in \mathcal{I}_{\text{test}}\},
\end{equation}
where $|\mathcal{I}_{\text{test}}| = N - L$. To ensure an unbiased evaluation, the same min-max normalization transformation learned from \( \mathcal{D}_{\text{train}} \cup \mathcal{D}_{\text{val}} \) is applied to \( \mathcal{D}_{\text{test}} \). This guarantees that the test set is never used in determining the scaling parameters, preventing data leakage and preserving the integrity of the evaluation.

Overall, the total dataset \( \mathcal{D}_{\text{tot}} \) contains \( N = 288 \) trajectories, obtained from 24 human participants (four per group across six groups), each performing two blocks of six trials. The training and validation data \( \mathcal{D} \) comprise \( L = 144 \) trajectories from the 12 participants in groups 1-3, following the same structure of two blocks and six trials per participant.

For training, we minimize the mean squared error loss using the Adam optimizer. Training is performed for 10 epochs with a mini-batch size of 32.

\subsection{Validation}
Validation is performed in an online manner, such that at each time step $t$, a sliding window of $w = 10$ of the most recent 3D position observations from the validation or test datasets is fed to the LSTM. This window continuously shifts forward as time progresses, ensuring the model always receives the latest $w=10$ steps to estimate the phase signal at the current time $t$. Additionally, since the LSTM model requires velocity signals, we compute them as discrete-time derivatives and concatenate the resulting velocities with the positions as input to our LSTM model.

For each trajectory $i$ in $\mathcal{D}_{\text{val}}$ or $\mathcal{D}_{\text{test}}$, we iteratively estimate the phase $\hat{\theta}^i_t$ for each time step $t$. By doing this for all time steps, the full predicted phase sequence $\hat{\boldsymbol{\Theta}}^{i} = [\hat{\theta}^{i}_1, \dots, \hat{\theta}^{i}_{T_i}]^\top$ is constructed for each $i \in \mathcal{I}_{\text{val}} \cup \mathcal{I}_{\text{test}}$. This sequence is then compared against the corresponding ground-truth phase sequence $\boldsymbol{\Theta}^{i}$ computed offline for all $i \in \mathcal{I}_{\text{val}} \cup \mathcal{I}_{\text{test}}$.

To evaluate the performance of the trained model, we define the circular phase error at each time step $t$ for trajectory $i$ as $\Delta \theta_t^{i} = \bigl|\arg\bigl(e^{j(\theta^{i}_t -\widehat{\theta}^{i}_t)}\bigr)\bigr|$. From these errors, we derive a performance metric expressed as the time average:
\begin{equation} \label{eq:global_metric}
\Delta \bar{\theta}^{i} = \frac{1}{|T_i|} \sum_{t=1}^{T_i} \Delta \theta^{i}_t,
\quad \forall i \in \mathcal{I}_{\text{val}} \cup \mathcal{I}_{\text{test}}.
\end{equation}

Furthermore, by averaging Eq.~\eqref{eq:global_metric} across all trajectories in the validation dataset and the test dataset, we obtain the overall average errors $\Delta\bar{\theta}_{\text{val}}$ and $\Delta\bar{\theta}_{\text{test}}$, respectively. The online nature of this validation procedure ensures that model performance is assessed under realistic conditions where phase estimates must be computed sequentially given only past observations of fixed lenght.

\begin{figure}[t]
\centering
\includegraphics[width=1\columnwidth]{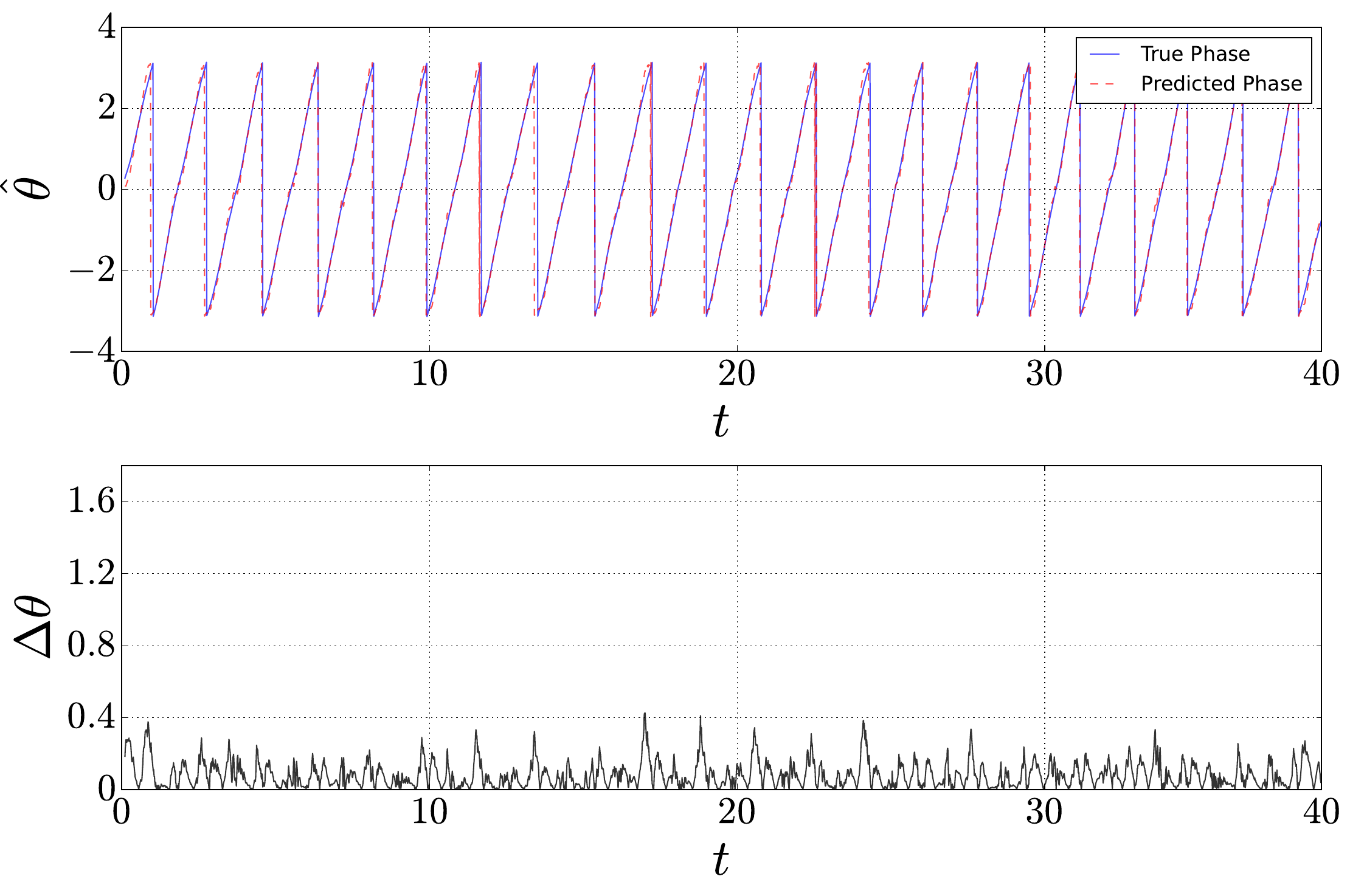} 
\caption{Comparison of the true phase (blue, solid line) of a trajectory from the test set $\mathcal{D}_{\text{test}}$ and the predicted phase (red, dashed line). The lower plot shows the circular error $\Delta \theta_t$.}
\label{fig:test_phases}
\end{figure}

Our simulation results demonstrate that the LSTM model achieves low errors on both validation (\(\Delta\bar{\theta}_\text{val}\) = $0.078\pm0.086\,\text{rad}$) and test dataset (\(\Delta\bar{\theta}_\text{test}\) = $0.158\pm0.110 \,\text{rad}$). The behavior of our online phase estimation is showed in Fig.~\ref{fig:test_phases}.

When deploying our phase estimator in real-world applications, sensor data might be collected in a global reference frame, making performing calibration over the entire trajectory as preprocessing infeasible. Therefore, we also evaluate the estimator's performance when calibration parameters are computed based on an initial segment of the trajectory before the online estimation begins.

Specifically, a buffer of $l$ samples is collected at the start of each test trajectory to compute the necessary calibration parameters using Eqs.~\eqref{eq:mean_calibration} and \eqref{eq:rotation_calibration}. This initial segment serves as a reference to estimate the required spatial transformations. Then, the online estimation proceeds on by applying the transformation to each point at each subsequent time step $t$, by the operations in Eqs.~\eqref{eq:shifted_traj} and \eqref{eq:rot_traj} the test set.

The results indicate that the error varies with the buffer size $l$. For instance, if \( l = 10\) (equivalent to 10 seconds), the average mean error is \( \Delta\bar{\theta}_\text{test}=0.267 \pm 0.110 \, \text{rad}\), while for \( l = 20\) (20 seconds), the error decreases to \( \Delta\bar{\theta}_\text{test}=0.178 \pm 0.109 \, \text{rad}\).

\section{Validation of LSTM Phase Estimation through RL agent}
\begin{figure*}[t]
\centering
\includegraphics[width=0.99\textwidth]{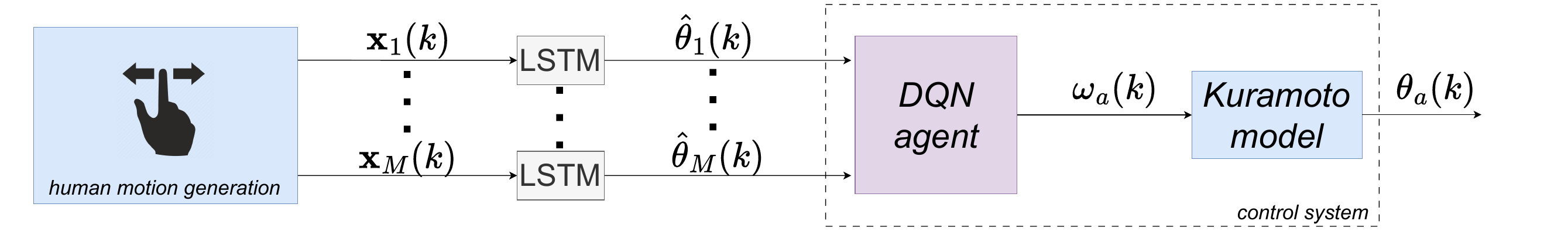}
\caption{Block diagram of the validation conducted using an RL agent. A synthetic dataset of 3D human motions is generated from pre-computed Kuramoto phases. Our LSTM model then extracts phase estimates online and feeds them to a pre-trained Deep Q-Network agent, which powers the control system (as in \cite{GROTTA202437}).}
\label{fig:dqn_estimation}
\end{figure*}

In the second validation stage, we assess the proposed phase estimation model using a dynamical system of coupled oscillators and a RL agent. This validation involves the following steps:

\begin{itemize}
\item \textbf{Replication of prior work:} We first replicate the methodology presented by Grotta et al. \cite{GROTTA202437}. This involves training a Deep Q-Network (DQN) agent (see \cite{Mnih2015DQN}) to achieve and/or enhance synchronization within a network of $M$ interacting Kuramoto oscillators \cite{kuramoto1975self}.
\item \textbf{Synthetic data generation:} The dynamics resulting from the synchronized Kuramoto simulation (specifically, the oscillator phase trajectories) are then used to generate a synthetic dataset of human-like 3D motions.
\item \textbf{Phase estimation:} We use our LSTM-based phase estimator to estimate the phase variable from this synthetic motion data.
\item \textbf{Evaluation:} Finally, we evaluate the impact of using the estimated phase on the synchronization task of the DQN agent. As illustrated in Fig.~\ref{fig:dqn_estimation}, the LSTM-estimated phase is provided as input to the DQN agent instead of the true phase from the simulation. We then measure the difference in synchronization performance achieved by the agent when using the estimated phases compared to when using the ground-truth phases.
\end{itemize}

To implement our validation protocol, we first simulate a network of coupled Kuramoto oscillators offline, where one of them is controlled by a DQN agent. The dynamics of the $i$-th oscillator (\(i=1,\dots,M\)) are governed by:
\begin{equation}
\dot{\theta}_i(t) = \omega_i(t) + c \sum_{j=1}^{M} A_{ij} \sin(\theta_j(t) - \theta_i(t)), \quad i=1,\dots, M
\label{eq:kuramoto}
\end{equation}
where \( \omega_i(t) \in \mathbb{R}_{>0} \) denotes the natural frequency of oscillator \( i \) at time \( t \), \( c \in \mathbb{R}_{\geq 0} \) represents the coupling strength between oscillators, and \( A_{ij} \) is the adjacency matrix defining the interaction topology. \( A_{ii} = 0 \) for all \( i \), while \( A_{ij} = 1 \) indicates that node \( j \) influences node \( i \), and \( A_{ij} = 0 \) otherwise. To enhance simulation realism, at each time step $t$, the natural frequencies $\omega_i(t)$ are drawn independently for each oscillator $i$ from a Gaussian distribution $\mathcal{N}(\mu_\omega, \sigma^2_\omega)$, while the coupling strength $c$ is kept constant at a value specified in \cite{alderisio2017a}.

The objective of the control agent is to maximize the coherence of the network's phase, quantified by the Kuramoto order parameter (see \cite{GROTTA202437, alderisio2017a}):
\begin{equation}
r(t) = \frac{1}{M} \left| \sum_{i=1}^{M} e^{j\theta_i(t)} \right|,
\end{equation}
A value of \( r(t) \approx 1 \) corresponds to near-perfect synchronization, whereas \( r(t) \approx 0 \) indicates incoherence. The system's long-term synchronization behavior is characterized by the time-averaged order parameter:
\begin{equation}
\langle r \rangle= \frac{1}{T} \int_{0}^{T} r(\tau) d\tau.
\label{eq:mean_order_parameter}
\end{equation}

The control system, depicted within the block diagram in Fig.~\ref{fig:dqn_estimation}, involves a Kuramoto model where the natural frequency $\omega_a$ of the controlled oscillator is adapted by a DQN controller to steer its dynamics within the network. The DQN agent receives state information derived from the oscillator phases $\theta_i(t)$ (obtained directly from the Kuramoto simulation in the baseline case) and outputs the desired natural frequency for the controlled oscillator. A more detailed explanation of the DQN agent's setup is provided in \cite{GROTTA202437}. In the \emph{human motion generation} block, we use the phase trajectories generated by the Kuramoto simulation (Eq.~\eqref{eq:kuramoto}) to create a synthetic dataset of 3D spatial positions.

Specifically, given the simulated phase trajectory \(\theta_i(t)\) for oscillator \( i \), a mapping function \(\mathcal{T}:[-\pi, \pi]\rightarrow\mathbb{R}^3\) is applied to transform phase values into spatial coordinates $\mathbf{x}_i(t) = [x_i(t), y_i(t), z_i(t)]^\top$. The transformation \(\mathcal{T}\) is designed to preserve the statistical properties of real human motion, ensuring that the generated trajectories are consistent with observed natural kinematics.

The transformation \(\mathcal{T}\) is defined as:
\begin{equation}
\mathcal{T}(\theta_i(t)) =
\begin{bmatrix}
\alpha_x +
\beta_x \cos(\theta_i(t)) + \mathcal{N}(0, \sigma) \\[8pt]
\alpha_y +
\beta_y \sin(\theta_i(t)) + \mathcal{N}(0, \sigma) \\[8pt]
\alpha_z + \mathcal{N}(0, \sigma)
\end{bmatrix},
\label{eq:3Dtransformation}
\end{equation}
where the parameters \( \alpha_x, \alpha_y, \alpha_z, \beta_x, \beta_y \) are computed from a single reference trajectory $\bar{\mathbf{X}} \in \mathbb{R}^{\bar{T} \times 3}$ from the test set \( \mathcal{D}_{\text{test}} \). The term \( \mathcal{N}(0, \sigma) \) represents Gaussian noise with mean \( 0 \) and standard deviation \( \sigma=0.1 \), added to introduce variability and better simulate real-world movement dynamics.

Specifically, $\alpha_x = \frac{1}{2}(\min(x) + \max(x))$ is the mid-range of the $x$-coordinates in the reference trajectory, defined as the average of its minimum ($\min(x)$) and maximum ($\max(x)$) values over time. This ensures the generated motion is centered around a realistic reference point within the range of human motion observed in \(\mathcal{D}_{\text{test}}\). The parameters $\alpha_y$ and $\alpha_z$ are defined similarly for the $y$ and $z$ coordinates.

The parameter $\beta_x = \frac{1}{2}(\max(x) - \min(x))$ determines the amplitude scaling of the oscillatory component along the \( x \)-axis, ensuring the synthesized trajectory remains within a plausible range of motion while preserving phase-dependent oscillatory behavior. Similarly, $\beta_y = \frac{1}{6}(\max(y) - \min(y))$ introduces controlled oscillations along the \( y \)-axis with a narrower scaling compared to the horizontal motion, replicating characteristics observed in the dataset where the $z$-coordinate displacement is kept constant.

Based on the cloesd-loop generated synthetic motion trajectories, we evaluate the control system’s performance under two conditions to assess the impact of using the estimated phases on synchronization performance, measured by the time-averaged order parameter $\langle r \rangle$ (Eq.~\eqref{eq:mean_order_parameter}):
\begin{itemize}
\item \emph{Without estimation}: The DQN uses the ground-truth phases $\theta_i(t)$ directly from the Kuramoto simulation.
\item \emph{With estimation}: The DQN agent uses phase values estimated in real-time by the LSTM network from the synthetic 3D position data.
\end{itemize}
\begin{table}[b]
\centering
\resizebox{0.49\textwidth}{!}{
\begin{tabular}{|c|c|c|}
\hline
Group & $\langle r \rangle$ (True Phase) & $\langle r \rangle$ (Estimated Phase) \\
\hline
1 & \( 0.7976 \pm 0.1020 \) & \( 0.7973\pm0.1024 \) \\
2 & \( 0.4956 \pm 0.1666 \) & \( 0.4962\pm0.1722 \) \\
\hline
\end{tabular}
}
\caption{Comparison of the time-averaged order parameter \( \langle r \rangle \) obtained using true phase values versus estimated phase values for Groups 1 and 2. A Student’s $t$-test for independent samples yielded \( p \)-values of \( 0.88 \) and \( 0.87 \) for Group 1 and 2, respectively, indicating no statistical significance.}
\label{tab:order_parameter_comparison}
\end{table}

Since RL can be sensitive to errors in state measurement, we tested the effect of the LSTM estimations on the final agent decisions across 5 independent trials. We used a calibration buffer of $l=3\Delta t$ (where $\Delta t$ is the simulation time step) and each trial was simulated for $30\,\text{s}$.

Numerical results are summarized in Table~\ref{tab:order_parameter_comparison} for two distinct groups considered in our study, each composed of $M=8$ oscillators, including one controlled agent. The mean and standard deviation of \( \langle r \rangle \) over the five trials for both cases (using true phases and estimated phases) are nearly identical, suggesting that the phase estimation method does not significantly degrade the agent's synchronization performance.

\begin{figure}
\centering
\includegraphics[width=1\columnwidth]{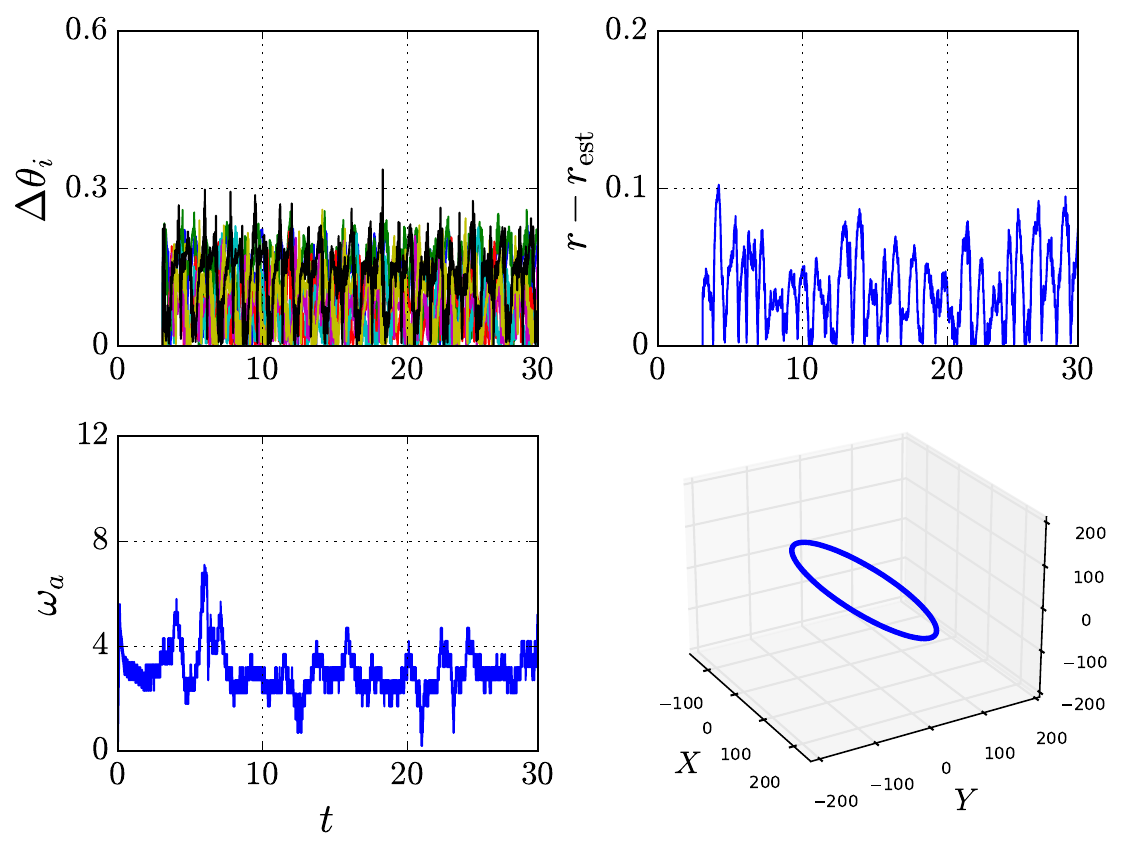}
\caption{Top left: Estimation error $\Delta \theta ^{i}$ for each oscillator over time.
Top right: Difference between the Kuramoto order parameter \( \langle r \rangle \) obtained with and without phase estimation.
Bottom left: Control input $\omega_a$ from the DQN agent.
Bottom right: Reconstructed 3D trajectory obtained using Eq.~\eqref{eq:3Dtransformation}.
Results refers to Group 2 as an exemplificative case.}
\label{fig:control_validation}
\end{figure}

Furthermore, we computed the average phase error across oscillators and over time, $\Delta\bar\theta=\frac{1}{M}\sum_{i=1}^M\Delta \bar\theta^{i}$, resulting in $\Delta\bar\theta = 0.0645 \pm 0.0220\,\text{rad}$ for Group 1 and $\Delta\bar\theta = 0.0922 \pm 0.0276\,\text{rad}$ for Group 2. Figure \ref{fig:control_validation} provides additional insights into the system's behavior for Group 2. The top left panel illustrates the estimation error $\Delta \theta ^{i}$ for each oscillator over time, while the top right panel shows the difference in the order parameter between using true and estimated phases. The bottom left panel depicts the control input $\omega_a$, while the bottom right panel displays the reconstructed 3D trajectory obtained using Eq.~\eqref{eq:3Dtransformation}.

\newpage
Eventually, we measured an execution time of the LSTM under 3 ms, and the combined execution time of the LSTM and DQN models of approximately 6 ms. These performance figures were obtained running on a MacBook Pro with an M1 processor and 8 GB of LPDDR4 RAM, demonstrating our proposed solution's suitability for real-time phase estimation compatible with modern 100 Hz motion capture systems.

\section{Conclusions}
This study presents a learning-based approach for estimating the phase of three-dimensional human motions, addressing limitations associated with traditional methods. By integrating an LSTM model with a spatial calibration procedure, the proposed framework ensures robust phase estimation that generalizes across different human subjects. The evaluation demonstrates that this approach provides low estimation errors in supervised learning experiments and maintains synchronization performance when applied within a reinforcement learning-based control system.

Our findings suggest that learning-based phase estimation can be effectively applied in real-world scenarios, particularly in applications involving human-avatar synchronization and interactive motion control. Future work will explore real-time implementation and integration with other systems or techniques to further enhance phase estimation capabilities in structured dynamic environments such as physiotherapy sessions.

\section{Acknowledgment}
The authors wish to acknowledge the contributions of Prof. Benoît G. Bardy and his team at University of Montpellier who provided the dataset used in this study, and Prof. Mario di Bernardo, University of Naples Federico II, for his insightful guidance during the development of this work.

\bibliographystyle{ieeetr} 

\bibliography{references}



\end{document}